# Generation and detection of Terahertz radiation by Field Effect Transistors

M. I. Dyakonov

Laboratoire Charles Coulomb, Université Montpellier2 - CNRS, France

This is an overview of the main physical ideas for application of field effect transistors for generation and detection of TeraHertz radiation. Resonant frequencies of the two-dimensional plasma oscillations in FETs increase with the reduction of the channel dimensions and reach the THz range for sub-micron gate lengths. When the mobility is high enough, the dynamics of a short channel FET at THz frequencies is dominated by plasma waves. This may result, on the one hand, in a spontaneous generation of plasma waves by a dc current and on the other hand, in a resonant response to the incoming radiation. In the opposite case, when plasma oscillations are overdamped, the FET can operate as an efficient broadband THz detector.

**1. Introduction**

The channel of a field effect transistor (FET) can act as a resonator for plasma waves with a typical wave velocity of $10^8$ cm/s. The plasma frequency of this resonator depends on its dimensions and for gate lengths of a micron and sub-micron size can reach the Terahertz (THz) range. The interest in the THz applications of FETs was initiated at the beginning of '90s by the theoretical work of Dyakonov and Shur [1] who predicted that a steady current flow in an asymmetric FET channel can lead to an instability against spontaneous generation of plasma waves. This will, in turn, produce the emission of electromagnetic radiation at the plasma wave frequency. Later, it was shown [2] that the nonlinear properties of the 2D plasma in the transistor channel can be used for detection and mixing of THz radiation. The resonant case of high electron mobility, when plasma oscillation modes are excited in the channel, and the non-resonant case of low mobility, where plasma oscillations are overdamped were analysed.

Both THz emission [3-6] and detection, resonant [7-9] and non-resonant [10-11], were observed experimentally at cryogenic, as well as at room temperatures, clearly demonstrating effects related to the excitation of plasma waves. At the moment, the most promising application appears to be the broadband THz detection and imaging in the overdamped regime, where plasma waves are non-existent. However, THz emission and resonant detection by excitation of plasma waves are also quite interesting phenomena that deserve further exploration.

**2. Plasma waves in low-dimensional structures**

Plasma waves are oscillations of the electron density. Generally, they can be obtained from the continuity equation:

$$\frac{\partial \rho}{\partial t} + \mathrm{div}\, \boldsymbol{j} = 0, \tag{1}$$

where $\rho$ is the charge density and $\boldsymbol{j}$ is the current density, related to the local electric field $\boldsymbol{E}$ by Ohm's law

$$\boldsymbol{j} = \sigma \boldsymbol{E}, \tag{2}$$

$\sigma$ is the conductivity. These equations must be complemented by the relation between the electric field and the charge density. In three dimensions, this relation obviously is div$E$ = $4\pi\rho/\varepsilon$, where $\varepsilon$ is the background dielectric constant (we use Gaussian units everywhere). In two and one dimensional structures, while this equation obviously remains true, it does not help because the field entering Eq. (2) is *not* the total electric field, but rather its component that can drive the current, e.g. for a two-dimensional electrons it is the part of electric field that lies in the 2D plane. (The div$j$ term in Eq. (1) should be also understood as divergence in two dimensions).

It should be taken into account that plasma waves exist in the high-frequency limit $\omega\tau > 1$, where $\omega$ is the frequency, and $\tau$ is the momentum relaxation time, which is also the damping time for plasma waves. Accordingly, if damping is completely ignored, we should use the high frequency limit for the complex conductivity: $\sigma(\omega) = ine^2/m\omega$, where $n$ is the electron concentration, $e$ and $m$ are the electron charge and effective mass respectively. This formula can be derived by writing the Drude equation for the mean electron velocity as $\partial v/\partial t = eE/m$ and neglecting the "friction" term $-v/\tau$. Equivalently, one can use the following equation for the current density $j = env$:

$$\frac{\partial j}{\partial t} = \frac{ne^2}{m} E .\qquad(3)$$

Combining Eqs. (1) and (3), we obtain

$$\frac{\partial^2 \rho}{\partial t^2} + \frac{ne^2}{m}\text{div}E = 0 .\qquad(4)$$

The electric field $E$ should be expressed through the charge density $\rho$, and this is where lies the difference between the 3D case and various low dimensional structures (gated or ungated 2D electrons, wires, etc).

**Bulk plasma waves.** For this case we have simply div$E$ = $4\pi\rho/\varepsilon$, and Eq. (4) describes a harmonic oscillator with a proper frequency

$$\omega_p = \sqrt{\frac{4\pi n e^2}{m\varepsilon}} ,\qquad(5)$$

which is the famous formula for the plasma frequency in three-dimensions (Langmuir waves).

**Two-dimensional electron gas.** We must express the electric field in the plane through the 2D charge density. This relation is given by an integral describing the Coulomb law. It is significantly simplified if one introduces the Fourier transforms for the charge density and the electric field, $\rho_k$ and $E_k$. Then it can be found that $E_k = -2\pi i k\rho_k/\varepsilon k$, (div$E)_k = 2\pi k\rho_k/\varepsilon$, and the Fourier transformed Eq. (5) becomes:

$$\frac{\partial^2 \rho_k}{\partial t^2} + \frac{2\pi n e^2 k}{m\varepsilon}\rho_k = 0 .\qquad(6)$$

Thus for a given wavevector $k$ the plasma wave frequency is given by the square root law:

$$\omega(k) = \sqrt{\frac{2\pi n e^2 k}{m\varepsilon}}, \tag{7}$$

where now *n* is the 2D electron concentration. Interestingly, a similar dispersion law $\omega \sim k^{1/2}$ describes surface waves in deep water (when the depth is much greater than the wavelength).

***Gated 2D electron gas.*** This is the case of a FET, which is the main object of our interest here. The relation between the charge density and the electric field in the channel is readily obtained from the plane capacitor formula:

$$\rho = en = CU, \tag{8}$$

where *C* is the gate-to-channel capacitance per unit area, and *U* is the so called gate voltage swing: $U = V_g - V_{th}$, $V_g$ is the gate voltage, and $V_{th}$ is the threshold voltage, at which the channel becomes completely depleted. From Eq. (8) we obtain:

$$\mathbf{E} = -\nabla U = -\frac{1}{C}\nabla\rho. \tag{9}$$

Note, that in contrast to the 3D case, where $\partial E_x/\partial x \sim \rho$, for gated 2D electrons we have $E_x \sim \partial\rho/\partial x$! It is important to understand that Eqs. (8, 9) hold not only when *U* is a constant, but also when the scale of the spatial variation of *U* is large compared to the gate-to-channel separation (the graduate channel approximation). Equation (4) now gives a linear dispersion relation for plasma waves:

$$\omega(k) = sk. \tag{10}$$

The plasma wave velocity *s* is given by:

$$s = \sqrt{\frac{ne^2}{mC}} = \sqrt{\frac{eU_0}{m}}. \tag{11}$$

where $U_0$ the dc part of the gate voltage swing related to the electron concentration *n* by Eq. (8).

It was shown in Ref. 1 that the nonlinear hydrodynamic equations describing the electrons in the channel of a FET are exactly the same as the shallow water equations in conventional hydrodynamics (the term "shallow water" refers to a situation when the wavelength, or more generally, the spatial scale of variation of the water level is much greater than the depth *h*). The only modification is that for the case of a FET one should replace *gh* (*g* is the free-fall acceleration) by *eU/m*. Thus plasma waves in the FET channel are analogous to shallow water waves, whose velocity is $(gh)^{1/2}$, compare to Eq. (11).

It should be reminded that the considerations in this subsection are based on Eq. (8), which is valid when the wavelength is much greater than the gate-to-channel separation *d*, or $kd \ll 1$. In the opposite case of short wavelengths ($kd \gg 1$) the existence of the gate is of no importance, and plasma waves are described by Eq. (7), similar to the "deep water" case.

Historically, plasma waves in two-dimensional structures were considered theoretically in Refs. 12-15, the first experimental observations were reported in Refs. 16, 17.

***Plasma waves in wires.*** The dispersion law for plasma waves is very similar to the previous case. The

Fourier transforms of the electrostatic potential $\varphi$ and the *linear* charge density $\rho$ are related by:

$$\varphi_k = (\rho_k / \varepsilon) \ln(1/kr). \qquad (12)$$

The wavelength is supposed to be large compared to the wire radius $r$ ($kr \ll 1$). Using this relation, and calculating $(\text{div}\mathbf{E})_k$ as above, one easily obtains from Eq. (4) a nearly linear dispersion law:

$$\omega(k) = sk \ln(1/kr), \quad s = \sqrt{\frac{ne^2}{m\varepsilon}}, \qquad (13)$$

where for this case $n$ is the 1D concentration and $\varepsilon$ is the dielectric constant of the surrounding medium.

### 3. Instability of the steady state with a dc current in FET

This instability was predicted in Ref. 1. The conditions for instability are:
 a) The plasma wave damping is small, $\omega\tau > 1$, where $\omega$ is the plasma oscillation frequency on the order of $s/L$, $L$ is the channel length, and $\tau$ is the momentum relaxation time defining the electron mobility.
 b) The boundary conditions at the source and the drain are asymmetric. An extreme case of such asymmetry, considered in Ref. 1, consists in the open circuit condition at the source and the short circuit condition at the drain.
 c) The steady state electron drift velocity $v$ must exceed a threshold value depending on the damping time $\tau$. For $\omega\tau \gg 1$, the threshold value of the drift velocity is *much smaller* than the plasma wave velocity $s$.
 The physical origin of this instability is related to the difference in velocities of plasma waves propagating upstream ($s-v$) and downstream ($s+v$). Because of this difference, the reflection coefficients at the boundaries may be *greater than 1*. It can be shown that for the boundary conditions mentioned above the net amplification due to plasma wave reflections during a round trip is equal to ($s+v$)/($s-v$). The time $t_0$ needed to make a round trip is obviously

$$t_0 = \frac{L}{s-v} + \frac{L}{s+v}. \qquad (14)$$

For $t \gg t_0$ the number of round trips can be estimated as $t/t_0$, and the total increase of the plasma wave amplitude during time $t$ can be written as $[(s+v)/(s-v)]^{t/t_0}$. We can now rewrite this expression as $\exp(\gamma t)$, where the instability increment $\gamma$ is given by the formula:

$$\gamma = \frac{s}{2L}\left(1 - \frac{v^2}{s^2}\right)\ln\left(\frac{s+v}{s-v}\right). \qquad (15)$$

This is the result obtained in Ref. 1 by the standard method of studying what happens to small perturbations of the steady state with a given drift velocity $v$. The dependence of the increment $\gamma$ on the ratio $v/s$ is presented in Fig. 1. For low drift velocities, $v \ll s$, Eq. (15) reduces to $\gamma = v/L$. In the absence of damping, the steady state is unstable for arbitrary small values of the drift velocity $v$, however if damping is taken into account, the instability occurs when $\gamma > 1/\tau$ and this condition defines the threshold value of the drift velocity. If $\tau$ is small enough, the steady state is stable.

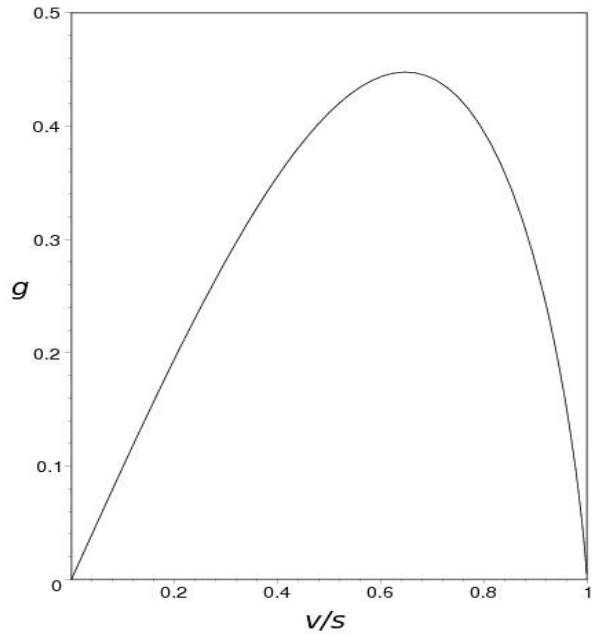

Fig. 1. The dimensionless instability increment $g = \gamma L/s$ as a function of $v/s$, Eq. (15)

The instability of the current-carrying steady state results in generation of plasma waves at the resonator modes. Essentially, the device operates like a laser, with an interesting difference: contrary to what happens in a laser, the gain is due to amplification *during reflections* from the "mirrors", while the losses occur during the propagation of the plasma wave between the mirrors.

The instability results in strong stationary nonlinear plasma oscillations in the channel. This was demonstrated in Ref. 18 by a numerical solution of the non-linearized "shallow water" equations. It was also shown that a similar instability may exist in an ungated two-dimensional electron gas [19].

The experimental results on THz emission from FETs in Refs. 3-6 and other work cannot be directly compared with the theory [1] because the experimental geometry is very different from the one-dimensional model adopted in [1]. In the standard experimental situation, the width $W$ of the gate is much larger than the gate length $L$, typically $W/L \sim 100$. Under such conditions the one-dimensional model, where the plasma density and velocity depend on the coordinate $x$ only, is not appropriate, since obviously oblique plasma waves with a non-zero component of the wave vector in the y direction can propagate. In such geometry, the gated region is not a resonator, but rather a waveguide with a continuous spectrum of plasma waves.

In Ref. 20, the analysis of stability was extended to the more realistic case when $W >> L$, and it was shown that, somewhat unexpectedly, in such a geometry an additional new mode of instability dominates, which is localized near the gate boundaries. Moreover, a similar instability should exist near a single boundary of current-carrying two-dimensional plasma.

Certainly, the linear theory cannot predict the outcome of this instability. However, since the spectrum of plasma waves is continuous, it seems likely that the instability will result in a turbulent motion of the electron fluid near the boundary of the gated region. The spectrum of the plasma oscillations should be broad, as it is observed in experiments. This is similar to what one can see in a river, when the water flows with sufficient velocity across an abrupt step in the waterbed: waves with wave vectors perpendicular to the flow are excited, while the wave vectors in the direction of the flow are purely imaginary, which accounts for the localization of the turbulent region near the step. It would be interesting to verify these predictions in specially designed experiments.

## 4. Detection of THz radiation by FET

The idea of using a FET for detection of THz radiation was put forward in Ref. 2. The possibility of the detection is due to nonlinear properties of the transistor, which lead to the rectification of an ac current induced by the incoming radiation. As a result, a photoresponse appears in the form of dc voltage between source and drain which is proportional to the radiation power (photovoltaic effect).

Obviously, some asymmetry between the source and drain is needed to induce such a voltage. There may be various reasons of such an asymmetry. One of them is the difference in the source and drain boundary conditions due to some external (parasitic) capacitances. Another one is the asymmetry in feeding the incoming radiation, which can be achieved either by using a special antenna, or by an asymmetric design of the source and drain contact pads. Thus the radiation may predominantly create an ac voltage between the source and the gate (or between the drain and the gate) pair of contacts. Finally, the asymmetry can naturally arise if a dc current is passed between source and drain, creating a depletion of the electron density on the drain side of the channel.

In most of the experiments carried out so far, the THz radiation was applied to the transistor channel, together with contact pads and bonding wires. In such a case, it is obviously difficult to define how exactly the radiation is coupled to the transistor. Theoretically, we will consider the case of an extreme asymmetry, where the incoming radiation creates an ac voltage with amplitude $U_a$ only between the source and the gate, see Fig. 2. We will also assume that there is no dc current between the source and drain.

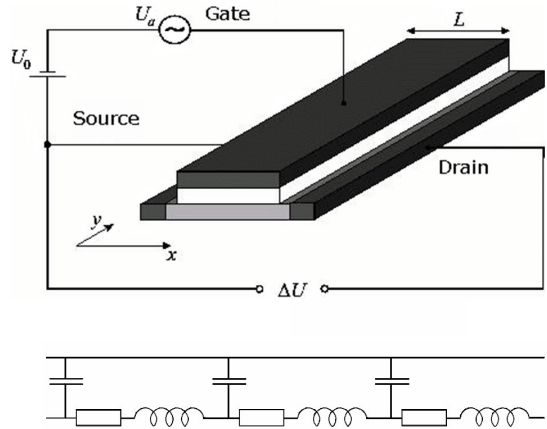

Fig. 2. Schematics of a FET as a THz detector (above) and the equivalent circuit (below).

Generally, the FET may be described by an equivalent circuit presented in Fig. 2. The obvious elements are the distributed gate-to-channel capacitance and the channel resistance, which depends on the gate voltage through the electron concentration in the channel, according to Eq. (8).

As mentioned above, this equation is valid locally, so long as the scale of the spatial variation of $U(x)$ is larger than the gate-to-channel separation $d$ (the gradual channel approximation). Under static conditions and in the absence of the drain current, $U = U_0 = V_g - V_{th}$, where $U_0$ is the static voltage swing. The inductances in Fig. 1 represent the so-called *kinetic* inductances, which are due to the electron inertia and are proportional to $m$, the electron effective mass. Depending on the frequency $\omega$, one can distinguish two regimes of operation, and each of them can be further divided into two sub-regimes depending on the gate length $L$.

**1. *High frequency regime*** occurs when $\omega\tau > 1$, where $\tau$ is the electron momentum relaxation time, determining the conductivity in the channel $\sigma = ne^2\tau/m$. In this case, the kinetic inductances in Fig. 2 are of primordial importance, and the plasma waves analogous to the waves in an RLC transmission line, will be excited. The plasma waves have a velocity $s = (eU_0/m)^{1/2}$ and a damping time $\tau$. Thus their propagation distance is $s\tau$.

**1*a*. *Short gate***, $L < s\tau$. The plasma wave reaches the drain side of the channel, gets reflected, and forms a standing wave with enhanced amplitude, so that the channel serves as a high-quality resonator for plasma oscillations. The fundamental mode has the frequency $\sim s/L$, with a numerical coefficient depending on the boundary conditions.

**1*b*. *Long gate***, $L >> s\tau$. The plasma waves excited at the source will decay before reaching the drain, so that the ac current will exist only in a small part of the channel adjacent to the source.

**2. *Low frequency regime***, $\omega\tau << 1$. Now, the plasma waves cannot exist because of overdamping. At these low frequencies, the inductance in Fig. 1 become simply short-circuits which leads to an RC line. Its properties further depend the gate length, the relevant parameter being $\omega\tau_{RC}$, where $\tau_{RC}$ is the RC time constant of the whole transistor. Since the total channel resistance is $L\rho/W$, and the total capacitance is $CWL$ (where $W$ is the gate width and $\rho = 1/\sigma$ is the channel resistivity), one finds $\tau_{RC} = L^2\rho C$.

**2*a*. *Short gate***, $L < (\rho C\omega)^{1/2}$. This means that $\omega\tau_{RC} < 1$, so that the ac current goes through the gate-to-channel capacitance practically uniformly on the whole length of the gate. This is the so-called "resistive mixer" regime [20-22]. For the THz frequencies this regime can apply only for transistors with extremely short gates smaller than 70 nm at 1 THz in silicon).

**2*b*. *Long gate***, $L >> (\rho C\omega)^{1/2}$. Now $\omega\tau_{RC} >> 1$, and the induced ac current will leak to the gate at a small distance $l$ from the source, such that the resistance $R(l)$ and the capacitance $C(l)$ of this piece of the transistor channel satisfy the condition $\omega\tau_{RC}(l) = 1$, where $\tau_{RC}(l) = R(l)C(l) = l^2\rho C$. This condition gives the value of the "leakage length" $l$ on the order of $(\rho C\omega)^{1/2}$ (which can also be rewritten as $s(\tau/\omega)^{1/2}$). If $l << L$, then neither ac voltage, nor ac current will exist in the channel at distances beyond $l$ from the source, see Fig. 3.

Thus, the characteristic length where the ac current exists is $s\tau$ for $\omega\tau > 1$, and $s(\tau/\omega)^{1/2}$ for $\omega\tau < 1$ [2]. Let us now present some quantitative examples for the different cases presented above. For $\tau = 30$ fs ($\mu = 300$ cm$^2$/(V·s) in Si MOSFET) and $s = 10^8$ cm/s the regime 1 will be realized for the radiation frequencies $f$ greater then 5 THz; the regime 1*a* for $L < 30$ nm. For $f = 0.5$ THz (regime 2), one finds the characteristic gate length distinguishing regimes 2*a* and 2*b* to be around 0.1 µm. If the conditions of the case 1*a* are satisfied, the photoresponse will be resonant, corresponding to the excitation of discrete plasma oscillation modes in the channel. Otherwise, the FET will operate as a broad-band detector.

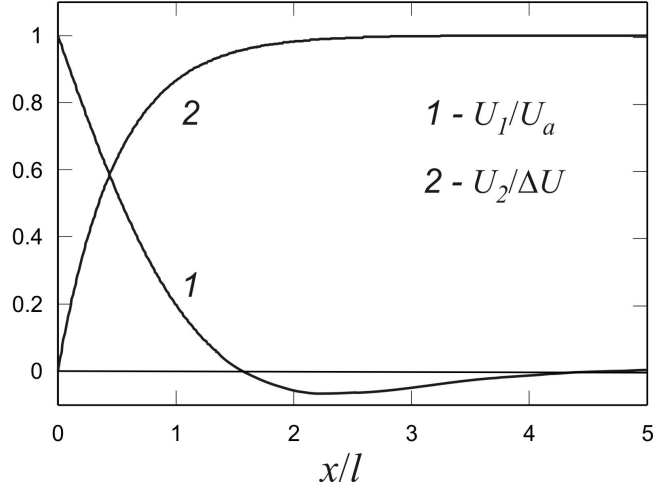

Fig. 3. Dependence of the ac voltage $U_1/U_a$ at $\omega t = 2\pi n$ and of the dc photoinduced voltage $U_2/\Delta U$ on the distance from the source $x$ for a long gate

For a long gate, there is no qualitative difference between the low-frequency regime ($\omega\tau \ll 1$), when plasma waves do not exist (the case 2b) and the high frequency regime ($\omega\tau \gg 1$), where plasma oscillations are excited (the case 1b). There is however some quantitative differences, see Eq. (16) below. Anyway, in the high frequency regime plasma waves are excited and their existence in the case 1b has been clearly confirmed by the recent detection experiments in magnetic field [21]. Plasma waves cannot propagate below the cyclotron frequency. Therefore, in experiments with a fixed radiation frequency the photoresponse is strongly reduced when the magnetic field goes through the cyclotron resonance [22]. This is probably the most spectacular manifestation of the importance of plasma waves in THz detection by FETs.

*Mechanism of the nonlinearity.* The most important mechanism is the modulation of the electron concentration in the channel, and hence of the channel resistance, by the local ac gate-to-channel voltage, as described by Eq. (8). Because of this, in the expression for the electric current $j = env$, both the concentration $n$, and the drift velocity $v$, will be modulated at the radiation frequency. As a result, a dc current will appear: $j_{dc} = e\langle n_1(t)v_1(t)\rangle$, where $n_1(t)$ and $v_1(t)$ are the modulated components of $n$ and $v$, and the angular brackets denote averaging over the oscillation period $2\pi/\omega$. Under open circuit conditions a compensating dc electric field will arise, resulting in the photoinduced source-drain voltage $\Delta U$.

*Simplified theory.* So far, the most important case is that of a long gate (the regimes 1b and 2b) when, independently of the value of the parameter $\omega\tau$, the ac current excited by the incoming radiation at the source cannot reach the drain side of the channel. For this case within the hydrodynamic approach the following result for the photoinduced voltage was derived [2]:

$$\Delta U = \frac{U_a^2}{4U_0}\left(1 + \frac{2\omega\tau}{\sqrt{1+(\omega\tau)^2}}\right). \tag{16}$$

As seen from this formula, the photoresponse changes only by a factor of 3, as the parameter $\omega\tau$ increases from low to high values, even though the physics becomes different: at $\omega\tau > 1$ plasma waves are excited, while at $\omega\tau < 1$ they are not. The basic equations may be written as [1, 2]:

$$\frac{\partial U}{\partial t} + \frac{\partial}{\partial x}(Uv) = 0, \qquad (17)$$

$$\frac{\partial v}{\partial t} = -\frac{e}{m}\frac{\partial U}{\partial x} - \frac{v}{\tau}. \qquad (18)$$

Here Eq. (17) is the continuity equation, in which the concentration $n$ is replaced by $U$ using Eq. 8, while Eq. (18) is the Drude equation for the drift velocity $v$ [23]. The boundary condition for gate-to-channel voltage at the source side of the channel ($x = 0$) is: $U(0,t) = U_0 + U_a \cos(\omega t)$. For a long gate, the boundary condition at the drain is $v(\infty) = 0$. The inertial term $\partial v/\partial t$ is accounted for by the kinetic inductances in Fig. 2. Here, we will consider only the simple case $\omega\tau < 1$, when the inertial term can be neglected. Then $v = -\mu \partial U/\partial x$, and

$$\frac{\partial U}{\partial t} = \mu \frac{\partial}{\partial x}\left(U\frac{\partial U}{\partial x}\right), \qquad (19)$$

where $\mu = e\tau/m$ is the electron mobility. We search the solution of the nonlinear Eq. (19) as an expansion in powers of $U_a$: $U = U_0 + U_1 + U_2$, $U_1$ is the ac voltage, proportional to $U_a$, and $U_2$ is the time-independent contribution proportional to $U_a^2$ (the photovoltage). In the first order in $U_a$ we obtain the diffusion equation for $U_1$ [24]:

$$\frac{\partial U_1}{\partial t} = s^2 \tau \frac{\partial^2 U_1}{\partial x^2}, \qquad (20)$$

with the boundary conditions $U_1(0,t) = U_a \cos(\omega t)$, $U_1(\infty,t) = 0$. The solution of this equation is

$$U_1(x,t) = U_a \exp(-x/l) \cos(\omega t - x/l), \qquad (21)$$

where the characteristic length $l$ for the decay of the ac voltage (and current) away from the source is given by:

$$l = s(2\tau/\omega)^{1/2}. \qquad (22)$$

This length defines the size of the part of the transistor adjacent to the source, whose resistance and the capacitance are such that $\omega\tau_{RC}(l) \sim 1$, as explained above.

In the second order in $U_a$, Eq. (5) yields:

$$U_0 \frac{\partial U_2}{\partial x} + \left\langle U_1 \frac{\partial U_1}{\partial x} \right\rangle = 0, \qquad (23)$$

which means simply the absence of the dc current. Integrating this equation, one obtains:

$$U_2(x) = \frac{1}{2U_0}\left[\left\langle U_1^2(0,t)\right\rangle - \left\langle U_1^2(x,t)\right\rangle\right], \qquad (24)$$

where the time averaged quantity $\left\langle U_1^2(x,t)\right\rangle = (1/2)U_a^2 \exp(-2x/l)$ is found from Eq. (21). Thus, the photovoltage $\Delta U = U_2(\infty)$ coincides with Eq. (16), provided that $\omega\tau \ll 1$. Figure 3 shows the ac voltage $U_1$ and the build-up of the dc voltage $U_2$ as functions of the distance from the source.

The maximal photovoltage is achieved at $U_0 \approx 0$, where the relative ac modulation of the electron concentration in the channel is the strongest (note that Eq. (8) is not valid in the near vicinity of $U_0 = 0$). A theoretical study of the photoresponse in this region is presented in Ref. 25.

It is instructive to compare the FET detector with the well known Schottky diode detector. In both devices the detection process is based on a rectification of the incident THz field by a nonlinear element. However, there are some important differences. The nonlinearity in the Schottky diode is due to the nonlinear I-V characteristic of the potential barrier between the metal and the semiconductor. The physical origin of the nonlinearity in the case of the FET transistor is very different. As discussed above, it is due to the fact that the incident THz radiation modulates both the carrier drift velocity and the carrier density. The static I-V dependence has no direct relevance to the detection properties of the FET.